\documentclass[aps,prd,12pt,a4paper,groupedaddress,preprintnumbers,floatfix,nofootinbib,showpacs]{revtex4}
\usepackage[english]{babel}
\usepackage{amsmath}
\usepackage{graphicx}
\usepackage{color}
\usepackage{amssymb}
\usepackage{hyperref}
\usepackage{dcolumn}
\usepackage{bm}
\usepackage{longtable}

\usepackage{appendix}
\usepackage{float}

\begin{document}
\preprint{DCP-12-04}
\title{Non-diagonal charged lepton mass matrix and non-zero $\theta_{13}$}
\author{J.~Alberto Acosta$^1$\footnote{Electronic address: kripton0x@gmail.com}, Alfredo Aranda$^{1,2}$\footnote{Electronic address: fefo@ucol.mx}, Manuel~A. Buen-Abad$^1$\footnote{Electronic address: manuelbuenabadnajar@gmail.com}, Alma D. Rojas$^1$\footnote{Electronic address: alma.drp@gmail.com}}

\affiliation{$^1$Facultad de Ciencias, CUICBAS,
Universidad de Colima, Colima, M\'exico \\
$^2$Dual C-P Institute of High Energy Physics, M\'exico }

\date{\today}

\begin{abstract}
Assuming that the neutrino mass matrix is diagonalized by the tribimaximal mixing matrix, we explore the textures for the charged lepton mass matrix that render 
an $U_{PMNS}$ lepton mixing matrix consistent with data. In particular we are interested in finding the textures with the maximum number of zeros. We explore
the cases of real matrices with three and four zeros and find that only ten matrices with three zeros provide solutions in agreement with data. We present the successful Yukawa textures including the relative sizes of their non-zero entries as well as some new and interesting relations among the entries of these textures in terms of the charged lepton masses. We also show that these relations can be obtained directly from a parametrization of the charged lepton mixing matrix $U_l$.
\end{abstract}

\pacs{11.30.Hv,	
12.15.Ff, 
14.60.Pq 
}

\maketitle

\section{Introduction}
In the framework of discrete flavor symmetries one of the most used ansatz  for the lepton 
mixing matrix $U_{PMNS}$ is the tribimaximal mixing (TBM) matrix~\cite{Harrison:2002er}, which despite the
non-zero value for $\theta_{13}$ recently confirmed by the T2K Collaboration~\cite{Abe:2011sj}, 
Double Chooz~\cite{Abe:2011fz}, Daya Bay~\cite{An:2012eh} and RENO~\cite{Ahn:2012nd} experiments, 
can still be used as a good approximation. Contributions from the charged lepton sector and/or from renormalization effects, can generate a non-zero
value for $\theta_{13}$ in agreement with the experimental data.    
For a classification of models predicting TBM mixing see~\cite{Barry:2010yk,Dorame:2011eb} and references therein, and for an overview
of flavor symmetry models for neutrino mixing, based on non-Abelian discrete symmetries that give the TBM pattern, see Ref.~\cite{Morisi:2012fg}.

Other ansatzs for the neutrino mass matrix have been explored. One such case consists on the so-called $n$-zero textures where, for example, taking the case of Majorana neutrinos, it has been found that only two independent zero entries are allowed by current data~\cite{Frampton:2002yf,Fritzsch:2011qv}. For Dirac mass matrices the situation is different (since the mass matrices are not symmetric) as
shown in~\cite{Hagedorn:2005kz}, where a study of the minimal mass matrices with as many zeros as possible determined that up to five zero entries are allowed. In both of these cases the neutrino mass matrices were expressed in the charged lepton diagonal basis. 
Fritzsch-like~\cite{Fritzsch} textures have also been considered for both the charged lepton and neutrino mass matrices in~\cite{Ahuja:2009jj}, where Dirac neutrinos are considered, and in~\cite{Mahajan:2010uv},
where all possible Hermitian six-zero\footnote{Here the counting of zeros refers to the total number of independent zeros in both the charged lepton and neutrino mass matrices.} Fritzsch-like as well as non Fritzsch-like textures for
lepton mass matrices have been investigated, for both Majorana and Dirac matrices. For a recent overview of all possible cases of Fritzsch-like as well as non Fritzsch-like matrices as possible textures of the fermion mass matrices see~\cite{Gupta:2011zzg}.

In this paper we take the following approach: assuming that the neutrino mass matrix is diagonalized by the TBM matrix, we determine the textures for the 
charged lepton mass matrix that leads to a lepton mixing matrix in agreement with
the experimental data. We find both the number and location of zeros in the charged lepton mass matrix as well as the relative sizes of the non-zero entries in the corresponding Yukawa matrix. The results obtained can be of use to model builders interested in obtaining flavor models with TBM mixing in the neutrino sector.

In Section~\ref{sec:setup} we present the general idea and setup of our analysis. The $n$-zero textures for $n=3$ are obtained in Section~\ref{sec:zeros}. Section~\ref{sec:analysis} contains a description of the numerical analysis performed, as well as our results and discussion. Finally, Section~\ref{sec:conclusions} contains our conclusions.

\section{The setup}\
\label{sec:setup}

The lepton mixing matrix $U_{PMNS}$ is given by
\begin{equation}\label{up}
 U_{PMNS}=U_{l}^{\dagger}U_{\nu},
\end{equation}
where $U_{l}$ relates left-handed leptons in the weak basis $e_{l}$ to the mass basis leptons $e'_{l}$:
$ e'_{l}=U_{l}e_{l}$; and $U_{\nu}$ corresponds to the same basis transformation for left-handed neutrinos $\nu'_{l}=U_{\nu}\nu_{l}.$

The $U_{PMNS}$ matrix can be parametrized as follows~\cite{PDG}:
\begin{equation}\label{upmns}
U_{PMNS} =\left(\begin{array}{ccc}
 c_{12}c_{13}&s_{12}c_{13}&s_{13}e^{-i\delta}\\
 -s_{12}c_{23}-c_{12}s_{23}s_{13}e^{i\delta}&c_{12}c_{23}-s_{12}s_{23}s_{13}e^{i\delta}&s_{23}c_{13}\\
 s_{12}s_{23}-c_{12}c_{23}s_{13}e^{i\delta}&-c_{12}s_{23}-s_{12}c_{23}s_{13}e^{i\delta}&c_{23}c_{13}
\end{array}\right)
\end{equation}
where $s_{ij}=\sin\theta_{ij}$, $c_{ij}=\cos\theta_{ij}$ and $\delta$ is the Dirac CP violating phase.

Since the $U$ matrices in \eqref{up} are unitary, we can express $U_{l}$ as

\begin{equation}\label{sl}
 U_{l}=U_{\nu}U_{PMNS}^{\dagger}.
\end{equation}

In our approach, we demand that $U_{\nu}$ be the TBM matrix, namely
\begin{equation}
U_{\nu}=U_{TBM} =\left(\begin{array}{ccc}
 \sqrt{\frac{2}{3}}&\frac{1}{\sqrt{3}}&0\\
 -\frac{1}{\sqrt{6}}&\frac{1}{\sqrt{3}}&-\frac{1}{\sqrt{2}}\\
-\frac{1}{\sqrt{6}}&\frac{1}{\sqrt{3}}&\frac{1}{\sqrt{2}}
\end{array}\right),
\end{equation}
and so, if $U_{\nu}=U_{TBM}$ is the matrix that diagonalizes the neutrino mass matrix
$M_{\nu}$ and $U_{l}$ the one that diagonalizes the {\it squared} charged lepton mass matrix $M^{2}_l \equiv M_{l}M^{\dagger}_{l}$, we
want to determine the form of $M_l$ such that $U_{PMNS}=U_{l}^{\dag}U_{TBM}$ has values in the allowed experimental range.

As described above the matrix $U_{l}$ diagonalizes $M^2_l$ and thus satisfies
\begin{equation}\label{equ}
 M^{2}_{l}=U_{l}M^{2}_{lD}U_{l}^{\dagger}
\end{equation}
where $M_{lD} = {\rm{diag}}(m_e,  \ m_{\mu},  \ m_{\tau})$ and $M^{2}_{lD} \equiv M_{lD}M^{\dagger}_{lD}$.

At this point $M_{l}$ is an arbitrary mass matrix with unknown values and we parametrize it as
\begin{equation}\label{gen}
M_{l}=\left(\begin{array}{ccc}
 a&b&c\\
 d&e&f\\
g&h&i
\end{array}\right).
\end{equation}
In general the mass matrix $M_{l}$ has complex entries, however, in this first approach we consider the particular case of a real mass matrix and thus assume all parameters in $M_l$ to be real. In this particular case, Eq.~\eqref{equ} becomes a set of equations for the entries of $M_{l}^{2}$:
\begin{equation}\label{mue}
 \left(\begin{array}{ccc}
 a^{2}+b^{2}+c^{2}&ad+be+cf&ag+bh+ci\\
 ad+be+cf&d^{2}+e^{2}+f^{2}&dg+eh+fi\\
ag+bh+ci&dg+eh+fi&g^{2}+h^{2}+i^{2}
\end{array}\right).
\end{equation}

There are nine variables and only six equations encoded in Eq.~\eqref{equ}. Thus we have an incomplete set of equations. To proceed we can start by making some assumptions on the {\it form} of the lepton mass matrix. Just as an example take $M_{l}$ to be a lower triangular matrix (by making $b=c=f=0$). This gives
\begin{equation}\label{muem}
 M_{l}M^{\dag}_{l}=\left(\begin{array}{ccc}
 a^{2}&ad&ag\\
 ad&d^{2}+e^{2}&dg+eh\\
ag&dg+eh&g^{2}+h^{2}+i^{2}
\end{array}\right).
\end{equation}
The number of variables has now been reduced to six variables, and with the six independent equations (as we show later in the paper) it is easy to find solutions for the parameters in Eq.~(\ref{muem}) in agreement with the experimental data.

We are interested in determining the textures with the maximum number of zeros. The main motivation for this is that having such textures can be useful to model builders interested in obtaining them using an underlying flavor symmetry. With this in mind, we explore the textures with the maximum number of
zeros such that Eq.~\eqref{equ} has real solutions in the allowed experimental range for the mixing angles. Note that our assumption of real mass matrices implies that in our case the CP violating phase satisfies $\delta = 0$ or $\pi$. We also assume that the Majorana CP violating phases are zero but note however, that since they
can be factorized from $U_{PMNS}$ as a diagonal matrix, their inclusion would not affect the magnitudes of the $U_{PMNS}$ entries.

\section{Finding the textures}
\label{sec:zeros}
We are concerned with the possible textures for the non-diagonal mass matrix for charged leptons (as we said
before, assuming that the neutrino mass matrix is diagonalized by the TBM  matrix). We assume that the entries are real and so $M_{l}M^{\dag}_l=M_lM_l^{T}$. Zeros in the entries of our matrices may be due to some underlying symmetry(ies) and because of that we will look for the textures with the highest possible number of zeros.

As discussed above, we have a system of six coupled, quadratic equations and (in general) nine variables. Thus there is a need for some extra constraints on the variables. For matrices with no zeros at all we need three of this extra constraints, for matrices with one zero we need only two. We begin our analysis with three-zero mass matrices requiring no extra constraints.

A combinatorial analysis shows that, for the three-zero matrices, we have $\binom{9}{3}=84$ ways to place the three zeros and thus $84$
different matrix forms. Nevertheless, it is well known that in forming the real product $M_lM_l^{T}$, any permutation of the columns of the
matrix involved does not change the resulting product matrix. Thus, this reduces the number of matrices to $20$. From these textures we discard those with a row or column made of zeros because they are singular. There are $4$ cases of this and we are left with only $16$ different textures. Another condition that has to be satisfied, in view of the non-zero values for the charged lepton masses, is 
$Tr\ (M_lM_l^{T})\neq 0$.

Denoting the possible mass matrices by $M_{nk}$, where indexes stand for the $k^{th}$ texture of the non-diagonalized charged lepton matrix with $n$ zeros, for $n=3$ we obtain:
\begin{eqnarray}\label{textuu} 
&M_{301}=\left(\begin{array}{ccc} 
0 & 0 & c \\
d & e & 0 \\
g & h & i 
\end{array} \right); \
M_{302}=\left(\begin{array}{ccc} 
0 & 0 & c \\
d & e & f \\
g & h & 0 
\end{array} \right);\
M_{303}=\left(\begin{array}{ccc} 
0 & b & c \\
d & 0 & 0 \\
g & h & i 
\end{array} \right);\
M_{304}=\left(\begin{array}{ccc} 
0 & b & c \\
d & 0 & f \\
g & h & 0 
\end{array} \right);\nonumber\\
&M_{305}=\left(\begin{array}{ccc} 
0 & b & c \\
d & e & f \\
g & 0 & 0 
\end{array} \right); \
M_{306}=\left(\begin{array}{ccc} 
a & b & c \\
0 & 0 & f \\
g & h & 0 
\end{array} \right); \
M_{307}=\left(\begin{array}{ccc} 
a & b & c \\
0 & e & f \\
g & 0 & 0 
\end{array} \right); \
M_{308}=\left(\begin{array}{ccc} 
0 & b & c \\
0 & e & f \\
g & 0 & i 
\end{array} \right); \nonumber\\
&M_{309}=\left(\begin{array}{ccc} 
0 & b & c \\
0 & 0 & f \\
g & h & i 
\end{array} \right); \
M_{310}=\left(\begin{array}{ccc} 
a & 0 & 0 \\
d & e & 0 \\
g & h & i 
\end{array} \right); \
M_{311}=\left(\begin{array}{ccc} 
0 & b & c \\
d & e & f \\
0 & 0 & i 
\end{array} \right); \
M_{312}=\left(\begin{array}{ccc} 
0 & b & c \\
d & 0 & f \\
0 & h & i 
\end{array} \right); \nonumber\\
&M_{313}=\left(\begin{array}{ccc} 
0 & 0 & c \\
d & e & f \\
0 & h & i 
\end{array} \right); \
M_{314}=\left(\begin{array}{ccc} 
a & b & c \\
0 & e & f \\
0 & 0 & i 
\end{array} \right);\
M_{315}=\left(\begin{array}{ccc} 
a & b & c \\
0 & 0 & f \\
0 & h & i 
\end{array} \right); \
M_{316}=\left(\begin{array}{ccc} 
a & 0 & c \\
0 & e & f \\
0 & h & i 
\end{array} \right). \nonumber\\ 
\end{eqnarray}
 
As none of the matrices shown above is obtained from another by means of permuting columns, these are all the possible three-zero textures to be analyzed.

\section{Analysis}
\label{sec:analysis}

We are interested in identifying the textures which provide solutions to Eq.~\eqref{equ}. In order to find which textures provide solutions we performed (for each texture) a scan over the whole allowed experimental range for the mixing angles  taking the values $\delta=0$ and $\delta=\pi$ for the CP-violating phase.

To carry out the numerical analysis we used the experimental data at 3$\sigma$ from the  global neutrino data analysis~\cite{Schwetz:2011zk} 
\begin{equation}\label{experim1}
 \begin{array}{|c|c|c|}
\hline
& \text{Best Fit Value}& 3\sigma \text{ range}\\
\hline
\sin^2 \theta_{12} & 0.312 & 0.27-0.36\\
\sin^2\theta_{23}& 0.52 & 0.39-0.64\\
\delta & -0.61\pi & 0-2\pi  \\
       & (-0.41\pi) &      \\
\hline
 \end{array}
\end{equation}
with  normal (inverted) hierarchy, and the recently Daya Bay results \cite{Ahn:2012nd} (confirmed at 5$\sigma$) 
\begin{equation}\label{experim2}
 \sin^2 2\theta_{13}=0.092\pm 0.017,\\
\end{equation}
which can be rewritten as $\ \sin^2 \theta_{13}=0.0235 \pm 0.0045$.

For the charged leptons masses we use the values given in~\cite{PDG}
\begin{eqnarray}
 m_{e}&=&0.510998910\pm0.000000013~\text{ MeV}\nonumber \ ,\\
m_{\mu}&=&105.658367\pm0.000004~\text{ MeV}\nonumber \ ,\\
m_{\tau}&=&1776.82\pm0.16~\text{ MeV} \ .
\end{eqnarray}

For instance by taking the central values of the mixing angles and those of the 
charged lepton masses (setting  $\delta=0$) we obtain
\begin{equation}
U_{PMNS} =\left(
\begin{array}{ccc}
 0.819631 & 0.551952 & 0.153476 \\
 -0.478787 & 0.512846 & 0.712567 \\
 0.314593 & -0.657524 & 0.684612
\end{array}
\right),
\end{equation}
and from Eq.~\eqref{sl} we get for $U_{l}$
\begin{equation}\label{slv}
U_{l}=\left(
\begin{array}{ccc}
 0.987895 & -0.0948358 & -0.122758 \\
 -0.124467 & -0.0123053 & -0.992147 \\
 0.0925805 & 0.995417 & -0.0239603
\end{array}
\right).
\end{equation}

If we use these values for the example of the lower triangular matrix presented in Section~\ref{sec:setup}, equating Eq.~\eqref{muem} to the explicit value of the right hand side of Eq.~\eqref{equ}, and using Eq.~\eqref{slv} and $M^{2}_{lD}$ evaluated at the central value for the lepton masses, we find:
\begin{equation}\label{exa}
 |M_{l}|=\left(\begin{array}{ccc}
 218.35&0&0\\
 1761.06&79.71&0\\
37.70&106.87&5.51
\end{array}\right),\ 
|Y_{l}|=\left(\begin{array}{ccc}
 8.8\times10^{-4}&0&0\\
 7.2\times10^{-3}&3.2\times10^{-4}&0\\
1.5\times10^{-4}&4.3\times10^{-4}&2.0\times10^{-5}
\end{array}\right);
\end{equation}\\
where $|M_{l}|$ is in MeV and the Yukawa matrix $Y_{l}$ shown above is obtained from:
\begin{equation}\label{overhiggs}
 Y_{l}=\frac{1}{\langle H \rangle}M_{l},
\end{equation}
being $\langle H \rangle \approx 246$ GeV the vacuum expectation value (VEV) of the Higgs field~\cite{PDG} (here we are assuming only one Higgs doublet contributes to the mass matrix. In general in this paper we assume either that, or if there are more SU(2) doublets involved, that they all acquire VEV's of $O(10^2)$~GeV).

To perform the complete scan we used a grid of dimensions $30 \times 30\times 30$ in the experimental range of the three mixing angles. For each combination of the mixing angles that provides solutions to Eq.~\eqref{equ}, we select only the real solutions. Once those are determined, we then also determine the order of magnitude of the Yukawa mass entries by finding their minimum and maximum values.

\subsection{Results for three-zero textures}
\label{subsec:results}
From all the possible three-zero textures (shown in Eq.~\eqref{textuu}) only the following provide (real) solutions:
$M_{304}$, $M_{308}$, $M_{309}$, 
$M_{310}$, $M_{311}$, $M_{312}$, $M_{313}$, $M_{314}$, $M_{315}$, and $M_{316}$.

For the textures $M_{301}$, $M_{302}$, $M_{303}$, $M_{305}$, $M_{306}$, and $M_{307}$ we did not find any solution. The reason is that all these textures lead to an extra constriction because the matrix product $M_lM_l^T$ has a zero
entry. Those entries are $(M_{301}M_{301}^T)_{12}$, $(M_{302}M_{302}^T)_{13}$, $(M_{303}M_{303}^T)_{12}$, $(M_{305}M_{305}^T)_{13}$, $(M_{306}M_{306}^T)_{23}$, and $(M_{307}M_{307}^T)_{23}$.

As mentioned above, once we have performed all calculations (including the normalization of the mass matrix by the VEV of the Higgs field) we find the maximum and minimum orders of magnitude of the entries for each texture of the Yukawa matrix. They are shown here (up to $O(1)$ coefficients): 

{\tiny
 \begin{equation*}
 \begin{array}{cll}
\mathbf{Y}& \qquad \qquad \qquad\qquad\mathbf{\delta}=\mathbf{0 }& \qquad\qquad \qquad\qquad\mathbf{\delta}=\mathbf{\pi} \\
\hline\\
  |Y_{304}|\sim&\left(
\begin{array}{ccc}
 0 & 10^{-7}-10^{-5} & 10^{-3}  \\
 10^{-6}-10^{-3} & 0 & 10^{-2}\\
10^{-6}-10^{-4} & 10^{-7}-10^{-4} & 0
\end{array}
\right);&\left(
\begin{array}{ccc}
 0 & 10^{-6}-10^{-4} & 10^{-4}-10^{-3}  \\
 10^{-5}-10^{-3} & 0 & 10^{-2}\\
 10^{-5}-10^{-4} & 10^{-6}-10^{-3} & 0
\end{array}
\right);\\
&&\\
|Y_{308}|\sim& \left(
\begin{array}{ccc}
 0 & 10^{-4}-10^{-3} & 10^{-9}-10^{-3} \\
 0 & 10^{-3}-10^{-2} & 10^{-7}-10^{-2} \\
10^{-5}& 0 & 10^{-4}-10^{-3}
\end{array}
\right);& \left(
\begin{array}{ccc}
 0 & 10^{-4}-10^{-3} & 10^{-9}-10^{-3} \\
 0 & 10^{-3}-10^{-2} & 10^{-7}-10^{-2} \\
10^{-5}& 0 & 10^{-4}-10^{-3}
\end{array}
\right);\\
&&\\
|Y_{309}|\sim&
 \left(
\begin{array}{ccc}
 0 & 10^{-5}-10^{-4}  & 10^{-4}-10^{-3} \\
 0 & 0 & 10^{-2} \\
10^{-5} & 10^{-4} & 10^{-9}-10^{-3} 
\end{array}
\right);
&
\left(
\begin{array}{ccc}
 0 & 10^{-5}-10^{-4}  & 10^{-4}-10^{-3} \\
 0 & 0 & 10^{-2} \\
10^{-5} & 10^{-4} & 10^{-8}-10^{-3} 
\end{array}
\right);\\
&&\\
|Y_{310}|\sim&
  \left(
\begin{array}{ccc}
 10^{-4}-10^{-3} & 0 & 0 \\
10^{-2} & 10^{-4}-10^{-3} & 0 \\
 10^{-9}-10^{-3} & 10^{-4}-10^{-3} & 10^{-5}
\end{array}
\right);&
\left(
\begin{array}{ccc}
 10^{-4}-10^{-3} & 0 & 0 \\
10^{-2} & 10^{-4}-10^{-3} & 0 \\
 10^{-9}-10^{-3} & 10^{-4}-10^{-3} & 10^{-5}
\end{array}
\right);\\
&&\\
|Y_{311}|\sim&
\left(
\begin{array}{ccc}
 0 & 10^{-4}-10^{-3} & 10^{-9}-10^{-3} \\
10^{-5} & 10^{-3}-10^{-2} & 10^{-7}-10^{-2} \\
 0 & 0 & 10^{-4}-10^{-3}
\end{array}
\right);&
\left(
\begin{array}{ccc}
 0 & 10^{-4}-10^{-3} & 10^{-9}-10^{-3} \\
10^{-5} & 10^{-3}-10^{-2} & 10^{-7}-10^{-2} \\
 0 & 0 &  10^{-4}-10^{-3}
\end{array}
\right);\\
&&\\
|Y_{312}|\sim&
\left(
\begin{array}{ccc}
 0 & 10^{-5}-10^{-4} & 10^{-4}-10^{-3} \\
 10^{-5} & 0 & 10^{-2} \\
 0 & 10^{-4} & 10^{-9}-10^{-3}
\end{array}
\right);&
\left(
\begin{array}{ccc}
 0 & 10^{-5}-10^{-4} & 10^{-4}-10^{-3} \\
 10^{-5} & 0 & 10^{-2} \\
 0 & 10^{-4} & 10^{-8}-10^{-3}
\end{array}
\right);\\
&&\\
|Y_{313}|\sim&
\left(
\begin{array}{ccc}
 0 & 0 & 10^{-4}-10^{-3} \\
10^{-5} & 10^{-4}-10^{-3} & 10^{-2} \\
 0 & 10^{-4}-10^{-3} & 10^{-9}-10^{-3}
\end{array}
\right);&
\left(
\begin{array}{ccc}
 0 & 0 & 10^{-4}-10^{-3} \\
10^{-5} & 10^{-4}-10^{-3} & 10^{-2} \\
 0 & 10^{-4}-10^{-3} & 10^{-9}-10^{-3}
\end{array}
\right);\\
&&\\
|Y_{314}|\sim&
\left(
\begin{array}{ccc}
 10^{-6} & 10^{-4}-10^{-3} & 10^{-9}-10^{-3} \\
 0 & 10^{-3}-10^{-2} & 10^{-7}-10^{-2} \\
 0 & 0 & 10^{-4}-10^{-3}
\end{array}
\right);&
\left(
\begin{array}{ccc}
 10^{-6} & 10^{-4}-10^{-3} & 10^{-9}-10^{-3} \\
 0 & 10^{-3}-10^{-2} & 10^{-7}-10^{-2} \\
 0 & 0 & 10^{-4}-10^{-3}
\end{array}
\right);\\
&&\\
|Y_{315}|\sim&
\left(
\begin{array}{ccc}
  10^{-6} & 10^{-5}-10^{-4} & 10^{-4}-10^{-3} \\
 0 & 0 & 10^{-2} \\
 0 & 10^{-4} & 10^{-9}-10^{-3}
\end{array}
\right);&
\left(
\begin{array}{ccc}
  10^{-6} & 10^{-5}-10^{-4} & 10^{-4}-10^{-3} \\
 0 & 0 & 10^{-2} \\
 0 & 10^{-4} & 10^{-8}-10^{-3}
\end{array}
\right);\\
&&\\
|Y_{316}|\sim&
\left(
\begin{array}{ccc}
10^{-6} & 0 &  10^{-4}-10^{-3} \\
 0 & 10^{-4}-10^{-3} & 10^{-2} \\
 0 & 10^{-4}-10^{-3} & 10^{-9}-10^{-3}
\end{array}
\right);&
\left(
\begin{array}{ccc}
10^{-6} & 0 &  10^{-4}-10^{-3} \\
 0 & 10^{-4}-10^{-3} & 10^{-2} \\
 0 & 10^{-4}-10^{-3} & 10^{-9}-10^{-3}
\end{array}
\right).\\
 \end{array}
\end{equation*}}

\subsection{Solution volume}
As part of the analysis we performed, we searched for the solution volume of each three-zero texture, that is, the set of points given by the three mixing angles that make possible to find real solutions to the entries of the $M_{l}$ matrix.

In the case of textures $M_{308}$ -- $M_{316}$ the solution volume fills the complete space given by the experimental intervals of the three mixing angles (the experimentally allowed parameter space). This happens for both $\delta=0$ and $\delta=\pi$.

\begin{figure}[h]
 \centering
 \includegraphics[width=9cm]{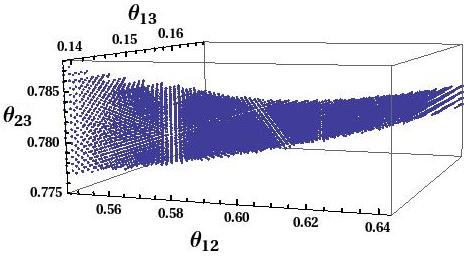}
 \caption{Solution volume of $M_{304}$ for $\delta=\pi$, found with a grid of $40\times 40 \times 40$. Similar results are obtained for $\delta=0$.}
 \label{fig1}
\end{figure}

The most interesting case is the texture $M_{304}$, which has the zeros on the major diagonal (and its permutations).
The relevance of this case is that the angle $\theta_{23}$ is now very restricted. Thus, we took a closer look within a smaller interval for $\theta_{23}$ using a finer grid of $40\times 40 \times 40$, and analyzing
a $\theta_{23}$ interval of $[0.7745,0.7950]$, we find that the solution volume of $M_{304}$, for both $\delta=0$ and $\delta=\pi$, 
is a thin curved surface (presented in Figure~\ref{fig1} for $\delta=\pi$); the only difference between these two cases is a 
displacement of the interval allowed for $\theta_{23}$, which is $[0.7763,0.7876]$ for $\delta=0$, and $[0.7750,0.7873]$ for $\delta=\pi$. 
These intervals are near but exclude the central value of $\theta_{23}$.

\subsection{A parametrization for $U_l$}

The Yukawa matrices in~\eqref{overhiggs} can explicitly be written as
\begin{equation}\label{ygen}
Y_{l}=\left(\begin{array}{ccc}
 y_a&y_b&y_c\\
 y_d&y_e&y_f\\
y_g&y_h&y_i
\end{array}\right).
\end{equation}
Then, the diagonal entries $(Y_{l}Y^{T}_{l})_{ii}$ are just the sum of the square of the elements of the i-th row of $Y_l$:

\begin{equation}\label{ySquareGen}
(Y_{l}Y_{l}^{T})_{11}=y_{a}^{2}+y_{b}^{2}+y_{c}^{2}, \quad (Y_{l}Y_{l}^{T})_{22}=y_{d}^{2}+y_{e}^{2}+y_{f}^{2},\quad (Y_{l}Y_{l}^{T})_{33}=y_{g}^{2}+y_{h}^{2}+y_{i}^{2}
\end{equation}

The textures obtained in Subsection~\ref{subsec:results} present some interesting features: first note that in all textures the largest upper bound ($10^{-2}$) always appears in the second row. Thus, from Eq.~\eqref{ySquareGen}, $(Y_{l}Y_{l}^{T})_{22} \sim O(10^{-4})$. This is a little higher than $O(m_{\tau}^2/{\langle H \rangle}^2 \sim 5\times10^{-5}$). Note also that (except for the $Y_{304}$) the third row has always an upper bound of $10^{-3}$ and then, from Eq.~\eqref{ySquareGen}, we can
deduce that $(Y_{l}Y_{l}^{T})_{33} \sim O(10^{-6})$. This is quite close to the value of the ratio
$m_{\mu}^2/{\langle H \rangle}^2 \sim 10^{-7}$.
More precisely and using the explicit values of the Yukawa matrix of Eq.~\eqref{exa} we find

\begin{eqnarray}\label{yuk}
 y^{2}_{a}+y^{2}_{b}+y^{2}_{c}&=&7.7\times10^{-7}\\\nonumber
  y^{2}_{d}+y^{2}_{e}+y^{2}_{f}&=&5.2\times10^{-5}\\\nonumber
   y^{2}_{g}+y^{2}_{h}+y^{2}_{i}&=&2.1\times10^{-7}\\\nonumber
\end{eqnarray}

We thus find that there are extra conditions on the entries of $Y_l$ in terms of the charged leptons masses. 
Because of equation (\ref{equ}), we can expect that these conditions may be due to an internal structure of the mixing matrix $U_{l}$ for the charged leptons.
In what follows we show that there is indeed a parametrization for this matrix that leads to specific relations between the entries in the Yukawa matrices and the charged lepton masses that reproduce the above results, including the small deviations in the orders of magnitude.

Motivated by the fact that the charged leptons are analogous to the down type quarks in terms of representations of the Standard Model gauge group, we consider a CKM-like parametrization for the $U_{l}$ mixing matrix in terms of the three angles $\theta_{12}^l$, $\theta_{13}^l$ and $\theta_{23}^l$, taking values between 
$0$ and $\pi/2$, and all phases set to zero in order to use real values for the $U_{l}$ matrix, as we have done in the whole analysis. Then, we can write:
\begin{equation}\label{para}
U_{l} =\left(\begin{array}{ccc}
 1&0&0\\
 0&\cos\theta^{l}_{23}&\sin\theta^{l}_{23}\\
0&-\sin\theta^{l}_{23}&\cos\theta^{l}_{23}
\end{array}\right)\left(\begin{array}{ccc}
 \cos\theta^{l}_{13}&0&\sin\theta^{l}_{13}\\
 0&1&0\\
-\sin\theta^{l}_{13}&0&\cos\theta^{l}_{13}
\end{array}\right)\left(\begin{array}{ccc}
 \cos\theta^{l}_{12}&\sin\theta^{l}_{12}&0\\
 -\sin\theta^{l}_{12}&\cos\theta^{l}_{12}&0\\
0&0&1
\end{array}\right)
\end{equation}

\begin{figure}[h]
 \centering
 \includegraphics[height=7cm]{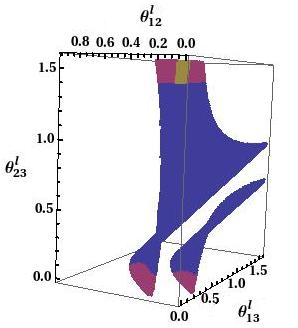}
 \caption{Sets of values for the angles of the charged lepton mixing matrix that match the experimental results (see description in text). The yellow (light) band in the upper part of the plot is the region consistent with all experimental results.}
 \label{fig2}
\end{figure}

Using $U'$ instead of $U_{PMNS}$ (to avoid agglomeration of indexes) one can easily find that:

\begin{equation}\label{e1}
 \sin^{2}\theta_{13}=1-(U'_{11})^{2}-(U'_{12})^{2} \ ,
\end{equation}

\begin{equation}\label{e2}
 \sin^{2}\theta_{23}=\frac{(U'_{23})^{2}}{(U'_{11})^{2}+(U'_{12})^{2}} \ ,
\end{equation}

\begin{equation}\label{e3}
 \sin^{2}\theta_{12}=\frac{(U'_{12})^{2}}{(U'_{11})^{2}+(U'_{12})^{2}} \ .
\end{equation}
 \\
Here the $\theta_{ij}$ are, as before, the parameters of $U_{PMNS}$; different from $\theta_{ij}^l$, which parametrize $U_l$.
We evaluated equations (\ref{e1})-(\ref{e3}) with $\theta^{l}_{ij}\in[0,\pi/2]$, and took all the points that matched the experimental intervals given by (\ref{experim1}) and (\ref{experim2}). The points ($\theta^{l}_{23},\theta^{l}_{13},\theta^{l}_{12}$) are presented in Figure~\ref{fig2}.

In Figure~\ref{fig2} the blue region is made of those values of the angles that solve equation (\ref{e1}) within the experimental values,
the purple points are those that solve equation (\ref{e1}) and (\ref{e2}) within the experimental ranges, and the dark yellow region is the set of points that satisfies the three equations (\ref{e1}), (\ref{e2}) and (\ref{e3}) within all the experimental values.
We then obtain a small region given by two small angles and a large one:

\begin{eqnarray}\label{values}
\theta^{l}_{12}&=&[0.06-0.15]\\
\theta^{l}_{13}&=&[0.07-0.16]\nonumber\\
\theta^{l}_{23}&=&[1.43-1.57\approx(\pi/2)]\nonumber
\end{eqnarray}

Now we use the parametrization of the $U_l$ matrix given by equation (\ref{para}) and the values of its angles to make 
an analysis of the entries in a general charged lepton mass matrix, of the form given by (\ref{gen}). Let us remember that $U_l$ appears in (\ref{equ}).
Using the values given in (\ref{values}), one can built a simpler form for $U_l$. Indeed, according
to equation (\ref{para}) and using the fact that $\theta^{l}_{12}$ and $\theta^{l}_{13}$ are small, we can use a first-order approximation
for their respective rotation matrices. The size of the angle $\theta^{l}_{23}$ allows us to use a first order approximation too, but around
$\theta^{l}_{23}=\pi/2$. Then, $\sin(\theta^{l}_{23})\approx 1$ and $\cos(\theta^{l}_{23})\approx \frac{\pi}{2}-\theta^{l}_{23} \equiv \epsilon$.
Thus, we end up with:

\begin{equation}\label{apro0}
U_{l} \approx\left(\begin{array}{ccc}
 1&0&0\\
 0&\epsilon&1\\
0&-1&\epsilon
\end{array}\right)\left(\begin{array}{ccc}
 1&0&\theta^{l}_{13}\\
 0&1&0\\
-\theta^{l}_{13}&0&1
\end{array}\right)\left(\begin{array}{ccc}
 1&\theta^{l}_{12}&0\\
 -\theta^{l}_{12}&1&0\\
0&0&1
\end{array}\right)
\end{equation}
 \\
Using (\ref{apro0}) in (\ref{equ}), we obtain the matrix $M^{2}_{l}$ in terms of $\theta_{12}^l$, $\theta_{13}^l$ and $\epsilon$.
Comparing the result with (\ref{mue}), we can see that the variables from the first row of the mass matrix must satisfy:

\begin{equation}\label{row1}
  a^{2}+b^{2}+c^{2} \approx  m_{e}^{2} + m_{\mu}^{2} \theta_{12}^{l2} + m_{\tau}^2\theta_{13}^{l2}
\end{equation}

the variables of the second row of (\ref{gen}) satisfy:

\begin{equation}\label{row2}
 d^{2}+e^{2}+f^{2}\approx m_{\tau}^2+m_{e}^2(-\epsilon\theta_{12}^l-\theta_{13}^l)^2+m_{\mu}^2(\epsilon-\theta_{12}^l\theta_{13}^l)^2 \sim m_{\tau}^2
\end{equation}

and the third row of (\ref{gen}) must have a contribution of the form

\begin{equation}\label{row3}
 g^{2}+h^{2}+i^{2}\approx m_{\tau}^2 \epsilon^2+m_{e}^2(\theta_{12}^l-\epsilon\theta_{13}^l)^2 +m_{\mu}^2(-1-\epsilon\theta_{12}^l\theta_{13}^l)^2 \sim m_{\mu}^2
\end{equation}

These expressions explain the observations considered at the beginning of the present subsection, including the corrections to the order of magnitude estimates, and represent the main result of this analysis.

Taking our specific example above, dividing equations (\ref{row1})-(\ref{row3}) by $\langle H\rangle^{2}$ as in (\ref{overhiggs}), and using the values $\theta_{12}^l=0.09$, $\theta_{13}^l=0.1215$, $\theta_{23}^l=1.55$ as well as the central values for the charged lepton masses we have:
\begin{eqnarray}
\frac{1}{\langle H\rangle^{2}}(m_{e}^{2} + m_{\mu}^{2} \theta_{12}^{l2} + m_{\tau}^2\theta_{13}^{l2})&=&7.7\times10^{-7}\\\nonumber
\frac{1}{\langle H\rangle^{2}}(m_{\tau}^2+m_{e}^2(-\epsilon\theta_{12}^l-\theta_{13}^l)^2+m_{\mu}^2(\epsilon-\theta_{12}^l\theta_{13}^l)^2)&=&5.2\times10^{-5}\\\nonumber
 \frac{1}{\langle H\rangle^{2}}(m_{\tau}^2 \epsilon^2+m_{e}^2(\theta_{12}^l-\epsilon\theta_{13}^l)^2 +m_{\mu}^2(-1-\epsilon\theta_{12}^l\theta_{13}^l)^2)&=&2.1\times10^{-7}\\\nonumber
\end{eqnarray}
These are the values from (\ref{yuk}), which are relations that exhibit the dependence of the entries of $Y_l$ on the masses and the mixing angles of $U_l$. It is easily observed that for the first row, although the leading term (zeroth order on the angles) is $m_{e}^2$, the other contributions
play a more important role than this one, because the masses of the $\mu$ and $\tau$ leptons are much greater than that of the electron and even overcome
the small size of their coefficients (which are powers of the angles). Finally, we note that these expressions are independent of the texture chosen
for the Yukawa matrix and also independent of the two values of the Dirac phase used in the PMNS matrix.

All the Yukawa textures found exhibit the relations given by equations (\ref{row1})-(\ref{row3}), so an important 
feature that we want to point out is that the matrices we have found show a very large mixing (nearly $\pi/2$) between the second and third 
generations of the charged leptons.

\subsection{Four-Zero textures}
We have already explored the textures with three zeros and found that it is possible to have solutions 
in the allowed ranges for many of them. Now we are interested in checking if this is also possible for 
textures with four zeros. To start, we list the possible textures with four zeros and their equivalent 
(column) permutations, discarding those with a row or column completely filled with zeros (thus 
avoiding singular matrices): 
\begin{eqnarray}
 \small
&M_{401}=\left(
\begin{array}{ccc}
 0 & 0 & c \\
 0 & e & 0 \\
 g & h & i
\end{array}
\right);\quad
M_{402}=\left(
\begin{array}{ccc}
 0 & 0 & c \\
 0 & e & f \\
 g & h & 0
\end{array}
\right);\quad
M_{403}=\left(
\begin{array}{ccc}
 0 & b & c \\
 0 & 0 & f \\
 g & h & 0
\end{array}
\right);\quad
M_{404}=\left(
\begin{array}{ccc}
 0 & b & c \\
 0 & e & f \\
 g & 0 & 0
\end{array}
\right);\nonumber\\
&M_{405}=\left(
\begin{array}{ccc}
 0 & 0 & c \\
 d & e & 0 \\
 0 & h & i
\end{array}
\right);\quad
M_{406}=\left(
\begin{array}{ccc}
 0 & 0 & c \\
 d & e & f \\
 0 & h & 0
\end{array}
\right);\quad
 M_{407}=\left(
\begin{array}{ccc}
 0 & 0 & c \\
 d & e & 0 \\
 g & h & 0
\end{array}
\right);\quad
 M_{408}=\left(
\begin{array}{ccc}
 0 & b & c \\
 d & 0 & 0 \\
 0 & h & i
\end{array}
\right);\nonumber\\
&M_{409}=\left(
\begin{array}{ccc}
 0 & b & c \\
 d & 0 & f \\
 0 & h & 0
\end{array}
\right);\quad
M_{410}= \left(
\begin{array}{ccc}
 a & b & c \\
 0 & 0 & f \\
 0 & h & 0
\end{array}
\right);\quad
M_{411}=\left(
\begin{array}{ccc}
 0 & 0 & c \\
 0 & e & f \\
 g & 0 & i
\end{array}
\right);\quad
M_{412}=\left(
\begin{array}{ccc}
 0 & b & 0 \\
 d & e & f \\
 0 & h & 0
\end{array}
\right);\nonumber\\
&M_{413}=\left(
\begin{array}{ccc}
 a & 0 & c \\
 0 & e & 0 \\
 0 & h & i
\end{array}
\right);\quad
M_{414}=\left(
\begin{array}{ccc}
 a & 0 & c \\
 0 & e & f \\
 0 & h & 0
\end{array}
\right);\quad
M_{415}=\left(
\begin{array}{ccc}
 0 & b & c \\
 0 & 0 & f \\
 g & 0 & i
\end{array}
\right);\quad
M_{416}=\left(
\begin{array}{ccc}
 a & b & c \\
 0 & 0 & f \\
 0 & 0 & i
\end{array}
\right);\nonumber\\
&M_{417}=\left(
\begin{array}{ccc}
 a & 0 & c \\
 0 & e & f \\
 0 & 0 & i
\end{array}
\right);\quad
M_{418}=\left(
\begin{array}{ccc}
 0 & 0 & c \\
 0 & 0 & f \\
 g & h & i
\end{array}
\right).
\end{eqnarray}

We have also calculated the product $M_lM_l^T$ for each texture and we can
observe that most of them render one or two zero entries. As in the previous case, this increases the difficulty in finding solutions to the system of equations in Eq.~\eqref{equ}. Moreover, since we are assuming only real values in the mass matrix entries, for the four-zero textures we are left with $6$ equations and only $5$ variables to satisfy them and the system is over-constrained.

The zero entries in the product  $M_lM_l^T$ for each texture are the following:
\begin{equation*}
 \begin{array}{ll}
(M_{401}M_{401}^T)_{12}=0,
&
(M_{402}M_{402}^{T})_{13}=0,\\
(M_{403}M_{403}^{T})_{23}=0,
&
  (M_{404}M_{404}^{T})_{13,23}=0,\\
(M_{405}M_{405}^{T})_{12}=0,
&(M_{406}M_{406}^{T})_{13}=0,\\
(M_{407}M_{407}^{T})_{12,13}=0,
&
(M_{408}M_{408}^{T})_{12}=0,\\
(M_{409}M_{409}^{T})_{23}=0,
&
(M_{410}M_{410}^{T})_{23}=
0,\\
(M_{413}M_{413}^{T})_{12}=0 &
(M_{414}M_{414}^{T})_{13}=0,\\
 \end{array}
\end{equation*}

The only textures which give non-zero entries in $M_lM_l^T$ are  $M_{411},M_{415},M_{416}$, and $M_{417} $. 
Those would be the more interesting cases 
to study if we allow complex parameters (this would imply also to allow the $\delta$ CP violating phase 
to take all its
possible values). $M_{412}$, $M_{416}$ and $M_{418}$ also render non-zero entries in $M_lM_l^{\dag}$ but 
they are singular.

We performed a scan over all the allowed experimental range for the mixing angles trying to
solve the system of equations for each particular texture, in the case of real parameters, without success. The only possibility is to consider cases where one of the zero entries in the four-zero texture gets slightly turned on. This case could correspond then to one of the three-zero texture above where one of the entries is much smaller than all others. Analyzing the three-zero textures and the relative size of their entries (presented above), we observe that except for $M_{304}$, all of them contain entries in the range $(10^{-4}-10^{-3})$ -- $10^{2}$~MeV, and that the smallest one is at least two orders of magnitude smaller than the next one. It is then possible, from a model building perspective, to consider a four-zero texture where one of the zeros is {\it lifted} (perhaps due to radiative corrections, or higher order operator mixings) such that at the end one ends up with one of the successful three-zero textures above. Here we show which four-zero texture could be used in such a scenario
\begin{equation*}
 \begin{array}{lllll}
M_{308} (c\sim 0)\rightarrow M_{402}; && M_{309} (i\sim 0)\rightarrow M_{403};&& M_{310} (g\sim 0)\rightarrow M_{402};\\
M_{311} (c\sim 0)\rightarrow M_{406}; && M_{312} (i\sim 0)\rightarrow M_{409};&& M_{313} (i\sim 0)\rightarrow M_{406};\\
M_{314} (c\sim 0)\rightarrow M_{414}; && M_{315} (i\sim 0)\rightarrow M_{410};&& M_{316} (i\sim 0)\rightarrow M_{414}.\\
 \end{array}
\end{equation*}

As a final remark we comment that even though at this point we considered only real matrices for the charged lepton mass matrices, it is possible to extend some of our results to the case of complex matrices with factorizable phases, i.e. mass matrices that can be written as $M_l=P^{*}\hat{M}_l P$, where $\hat{M}_l$ are matrices with real entries and  $P$ a diagonal phase matrix, $P=\text{diag}(e^{\imath\phi1},e^{\imath\phi2},e^{\imath\phi3})$ . Then, if $\mathcal{O}_l$ is the orthogonal matrix which diagonalizes $\hat{M}_l\hat{M}_l^{T}$ we can write, under our assumption $U_{\nu}=U_{TBM} $, 
$\mathcal{O}_l=  P^{*} U_{TBM} U_{PMNS}^T$. Nevertheless, from the property of orthogonality and $\det \mathcal{O}_l=\pm 1$ we can infer that the only possible values for the factorizable phases are integer multiples of $\pi$. 
We checked exactly the same three-zero textures as before, but allowing complex entries and a CP-violating phase $\delta=\pi/2$ and found solutions for the same textures as in the real case. The difference with our previous results is that in these cases solutions are found in all the allowed volume of angles, even for $Y_{304}$, and the order of magnitude of the entries are generally larger. In the four-zero texture we again find no solutions and due to the fact that the entries in the three-zero textures are larger in this case, it is not possible to easily start from a four-zero texture and lift one zero (here we are thinking of a naturally small correction to the zero entry in question).

\section{Conclusions}
\label{sec:conclusions}
We analyze several textures for the charged lepton mass matrix under the assumption that the neutrino mass matrix is diagonalized by the TBM matrix, and with the intention of maximizing the number of zeros in them. Requiring the $U_{PMNS}$ mixing matrix to be consistent with
the allowed experimental range for mixing angles (fixing the CP violating phase), and with (the central values of) the charged lepton masses, we explore the maximum number of zeros possible in a texture under the assumption of real charged lepton mass matrix entries. We find that there are ten three-zero textures which provide $U_{PMNS}$ values in agreement with and also determine the size range for their entries. Among the successful textures, the one with zeros in the diagonal shows an interesting behaviour in the sense that in order to work, it requires the mixing angle $\theta_{23}$ to lie in a very restricted range which, albeit consistent with the experimental range, it excludes the central experimental values.

A general analysis of the successful textures showed that there are relations between their entries and the charged lepton masses. Through a CKM-like parametrization of the $U_l$ mixing matrix we are able to obtain the texture-independent specific relations in terms of the three rotation angles in $U_l$.

We find no solutions for four-zero textures but observe that it is possible to consider cases where
one the zeros is lifted in such a way that the remaining mass matrix corresponds to some of the successful three-zero textures. Finally, we explore a possible extension to consider the cases with complex factorizable textures. We find similar results, i.e. solutions available for the three-zero textures and no solutions for the four-zero-texture, with the difference that the typical entry size of the charged lepton mass matrix is generally larger in this case.

While preparing this paper, a new preprint appeared in the arxiv~\cite{newpaper}, where an analysis of the charged lepton mixing matrix was performed using the TBM matrix form as well as the Bimaximal Mixing matrix (BM) for the neutrino mixing matrix. They consider the charged lepton mixing matrix to be nearly the identity but deviated from it by one and two parameters for the BM and the TBM case respectively. The parameters are the Cabbibo angle $\lambda=\sin \theta_{12}$ and $\sin\theta^l_{23}$ for the TBM case, assuming that the $U_{l}$ has its own CKM-like parametrization.  

In our case the assumption of $U_{l}$ as a matrix near the identity is not used. Instead, we determine this matrix by allowing the three rotation angles to vary over the whole experimental range and find that in fact, since $\theta^l_{23} \sim \pi/2$, the matrix is not close to the identity 

Also, in our context, the goal included to find the textures that are diagonalized with Eq.~\eqref{sl}. This is not explored in~~\cite{newpaper} and we feel it is a useful result that can be used by model builders interested in flavor symmetries.

\begin{acknowledgments}
M.~B-A. thanks conacyt for a fellowship as a SNI research assistant. A.A. acknowledges support from conacyt (M\'exico).

\end{acknowledgments}


\begin{thebibliography}{}
\bibitem{Harrison:2002er}
  P.~F.~Harrison, D.~H.~Perkins, W.~G.~Scott,
  ``Tri-bimaximal mixing and the neutrino oscillation data,''
  Phys.\ Lett.\  {\bf B530}, 167 (2002).
  [hep-ph/0202074].

\bibitem{Abe:2011sj}
  K.~Abe {\it et al.}  [T2K Collaboration],
  ``Indication of Electron Neutrino Appearance from an Accelerator-produced
  Off-axis Muon Neutrino Beam,''
  Phys.\ Rev.\ Lett.\  {\bf 107}, 041801 (2011)
  [arXiv:1106.2822 [hep-ex]].


\bibitem{Abe:2011fz}
  Y.~Abe {\it et al.},
  ``Indication for the disappearance of reactor electron antineutrinos in the
  Double Chooz experiment,''
  arXiv:1112.6353 [hep-ex].


\bibitem{An:2012eh} 
  F.~P.~An {\it et al.}  [DAYA-BAY Collaboration],
  ``Observation of electron-antineutrino disappearance at Daya Bay,''
  Phys.\ Rev.\ Lett.\  {\bf 108}, 171803 (2012)
  [arXiv:1203.1669 [hep-ex]].


\bibitem{Ahn:2012nd} 
  J.~K.~Ahn {\it et al.}  [RENO Collaboration],
  ``Observation of Reactor Electron Antineutrino Disappearance in the RENO Experiment,''
  Phys.\ Rev.\ Lett.\  {\bf 108}, 191802 (2012)
  [arXiv:1204.0626 [hep-ex]].


\bibitem{Barry:2010yk} 
  J.~Barry and W.~Rodejohann,
  ``Neutrino Mass Sum-rules in Flavor Symmetry Models,''
  Nucl.\ Phys.\ B {\bf 842}, 33 (2011)
  [arXiv:1007.5217 [hep-ph]].


\bibitem{Dorame:2011eb}
L.~Dorame, D.~Meloni, S.~Morisi, E.~Peinado and J.~W.~F.~Valle,
  ``Constraining neutrinoless double beta decay,''
  Nucl.\ Phys.\ B {\bf 861}, 259 (2012)
  [arXiv:1111.5614 [hep-ph]].

\bibitem{Morisi:2012fg} 
  S.~Morisi and J.~W.~F.~Valle,
  ``Neutrino masses and mixing: a flavour symmetry roadmap,''
  arXiv:1206.6678 [hep-ph].


\bibitem{Frampton:2002yf} 
  P.~H.~Frampton, S.~L.~Glashow and D.~Marfatia,
  ``Zeroes of the neutrino mass matrix,''
  Phys.\ Lett.\ B {\bf 536}, 79 (2002)
  [hep-ph/0201008].

\bibitem{Fritzsch:2011qv} 
  H.~Fritzsch, Z.~-z.~Xing and S.~Zhou,
  ``Two-zero Textures of the Majorana Neutrino Mass Matrix and Current Experimental Tests,''
  JHEP {\bf 1109}, 083 (2011)
  [arXiv:1108.4534 [hep-ph]].

\bibitem{Hagedorn:2005kz} 
  C.~Hagedorn and W.~Rodejohann,
  ``Minimal mass matrices for dirac neutrinos,''
  JHEP {\bf 0507}, 034 (2005)
  [hep-ph/0503143].

\bibitem{Fritzsch}
  H.~Fritzsch, 
  Phys.\ Lett.\ B {\bf 70}, 436 (1977); ibid.  ``Weak Interaction Mixing in the Six - Quark Theory,''
  Phys.\ Lett.\ B {\bf 73}, 317 (1978).


\bibitem{Ahuja:2009jj} 
  G.~Ahuja, M.~Gupta, M.~Randhawa and R.~Verma,
  ``Texture specific mass matrices with Dirac neutrinos and their implications,''
  Phys.\ Rev.\ D {\bf 79}, 093006 (2009)
  [arXiv:0904.4534 [hep-ph]].

\bibitem{Mahajan:2010uv} 
  N.~Mahajan, M.~Randhawa, M.~Gupta and P.~S.~Gill,
  ``Investigating texture six zero lepton mass matrices,''
  arXiv:1010.5640 [hep-ph].

\bibitem{Gupta:2011zzg} 
  M.~Gupta and G.~Ahuja,
  ``Possible textures of the fermion mass matrices,''
  Int.\ J.\ Mod.\ Phys.\ A {\bf 26}, 2973 (2011)
  [arXiv:1206.3844 [hep-ph]].

\bibitem{PDG}
  K.~Nakamura {\it et al.} [ Particle Data Group Collaboration ],
  ``Review of particle physics,''
  J.\ Phys.\ G {\bf G37}, 075021 (2010) and 2011 partial update for the 2012 edition (URL: http://pdg.lbl.gov).


\bibitem{Schwetz:2011zk}
  T.~Schwetz, M.~Tortola, J.~W.~F.~Valle,
  ``Where we are on $\theta_{13}$: addendum to 'Global neutrino data and recent reactor fluxes: status of three-flavour oscillation parameters',''
  New J.\ Phys.\  {\bf 13}, 109401 (2011).
  [arXiv:1108.1376 [hep-ph]].

\bibitem{newpaper}
  C.~Duarah, A.~Das, N.~Nimai Singh,
  ``Charged lepton contributions to bimaximal and tri-bimaximal mixings for generating $\sin\theta_{13}\neq 0$ and $\tan^2\theta_{23}<1$,''
  [hep-ph/1207.5225v1].


\end{thebibliography}
\end{document}